# Investigation of the temperature-dependence of ferromagnetic resonance and spin waves in $Co_2FeAl_{0.5}Si_{0.5}$

Li Ming Loong[1], Jae Hyun Kwon[1], Praveen Deorani[1], Chris Nga Tung Yu[2], Atsufumi Hirohata[3,a], and Hyunsoo Yang[1,b]

[1]Department of Electrical and Computer Engineering, National University of Singapore, 117576 Singapore
[2]Department of Physics, The University of York, York, YO10 5DD, UK
[3]Department of Electronics, The University of York, York, YO10 5DD, UK

$Co_2FeAl_{0.5}Si_{0.5}$ (CFAS) is a Heusler compound that is of interest for spintronics applications, due to its high spin polarization and relatively low Gilbert damping constant. In this study, the behavior of ferromagnetic resonance as a function of temperature was investigated in CFAS, yielding a decreasing trend of damping constant as the temperature was increased from 13 to 300 K. Furthermore, we studied spin waves in CFAS using both frequency domain and time domain techniques, obtaining group velocities and attenuation lengths as high as 26 km/s and 23.3 μm, respectively, at room temperature.

[a] Electronic mail: atsufumi.hirohata@york.ac.uk
[b] Electronic mail: eleyang@nus.edu.sg

Half-metallic Heusler compounds with low Gilbert damping constant (α) are promising candidates for spin transfer torque-based (STT) spintronic devices,[1-3] spin-based logic systems,[4] as well as spin wave-based data communication in microelectronic circuits.[5] Hence, a deeper fundamental understanding of the magnetization dynamics, such as the behavior of ferromagnetic resonance (FMR) and spin waves in Heusler compounds, could enable better engineering and utilization of these compounds for the aforementioned applications. In previous work, FMR has been investigated in several Heusler compounds, such as $Co_2FeAl$ (CFA),[6] $Co_2MnSi$ (CMS),[7] and $Co_2FeAl_{0.5}Si_{0.5}$ (CFAS).[8] In addition, the variation of α with temperature has been studied for other materials, such as Co, Fe, Ni, and CoFeB.[9-11] However, the temperature-dependence of α in Heusler compounds has not been reported yet. Furthermore, while there have been some studies of spin waves in Heusler compounds, such as CMS and $Co_2Mn_{0.6}Fe_{0.4}Si$ (CMFS),[7,12] these studies have focused on frequency domain measurements. Thus, time domain measurements remain scarce, and mainly consist of time-resolved magneto-optic Kerr effect (TR-MOKE) experiments.[13] In this work, we investigate the temperature-dependence of α in CFAS, a half-metallic Heusler compound.[14,15] Moreover, we utilize both frequency domain and pulsed inductive microwave magnetometry (PIMM) time domain measurements to study the magnetization dynamics in CFAS. We obtain α of 0.0025 at room temperature, which is 6 times lower than the value at 13 K. In addition, we evaluate the group velocity ($v_g$) and the attenuation length (Λ) in CFAS, leading to values as high as 26 km/s and 23.3 μm respectively, at room temperature.

CFAS (30 nm thick) was grown by ultrahigh vacuum (UHV) molecular beam epitaxy (MBE) on single crystal MgO (001) substrates and capped with 5 nm of Au. The base pressure was $1.2 \times 10^{-8}$ Pa and the pressure during deposition was typically $1.6 \times 10^{-7}$ Pa. The substrates were cleaned with acetone, IPA and deionised water in an ultrasonic bath before being loaded into the chamber. After the film growth, the samples were *in-situ* annealed at 600 °C for 1 hour. CFAS alloy and Au pellets were used as targets for electron-beam bombardment. Figure 1(a) shows the vibrating sample magnetometry (VSM) results, from which the saturation magnetization ($M_s$) was extracted. The measurement was also repeated at different temperatures to extract the corresponding values of $M_s$ for subsequent data fitting. The $M_s$ value increases from 1100 emu/cc at 300 K, to 1160 emu/cc at 13 K. From the VSM data, we verify a hard axis along [100] and an easy axis along [110], consistent with earlier reports.[3,8] In addition, the θ-2θ XRD data shown in Fig. 1(b) verified the presence of the characteristic (004) peak, indicating that the CFAS film was at least B2-ordered.[1,14] As shown in Fig. 1(c), the film was patterned into mesas, which were integrated with asymmetric coplanar waveguides (ACPW). The ACPWs were electrically isolated from the mesa by 50 nm of $Al_2O_3$, which was deposited by RF sputtering. Vector network analyzer (VNA) and PIMM techniques were used to excite and detect ferromagnetic resonance (FMR) as well as spin waves in CFAS. The former technique allows frequency domain measurements, while the latter technique was used for time domain measurements. The experimental setup enabled the excitation of Damon-Eshbach-type (DE) modes, as the external magnetic field was applied along the ACPWs, shown in Fig. 1(c).[16]

A VNA was connected to the ACPWs, and reflection as well as transmission signals were measured to study the FMR and spin wave propagation, respectively. Background subtraction was performed to obtain the resonance peaks. Figure 2(a) shows the FMR frequency as a function of applied magnetic field at different temperatures, with the corresponding fits using the Kittel formula,[17]

$$f = \frac{\gamma}{2\pi}\sqrt{(H + H_a)(H + H_a + 4\pi M_s)}, \qquad (1)$$

where $f$ is the resonance frequency, $\gamma$ is the gyromagnetic ratio, $H$ is the applied magnetic field, and $H_a$ is the anisotropy field. The $g$ factor, which was extracted using the equation $\gamma = 2\pi g \mu_B/h$, where $\mu_B$ is the Bohr magneton and $h$ is Planck's constant, was found to be 2.03±0.02, while $H_a$ generally decreased from 130 Oe at 13 K to 70 Oe at 300 K. The $(g – 2)$ value is lower than those of Co and Ni, but comparable to those of other Heusler compounds, such as CMS and $Co_2MnAl$ (CMA).[18] The deviation of the $g$ factor from the free electron value of 2 is correlated with the spin-orbit interaction in a material, where a smaller deviation indicates weaker spin-orbit interaction, and lower $\alpha$.[18] The inset of Fig. 2(a) shows the resonance frequency at $H = 1040$ Oe as a function of temperature, with a Bloch fitting, indicating a Curie temperature of approximately 1000 K. The Bloch fitting was performed by substituting the following equation[19] into Eq. (1):

$$M_s = a_0\left(1 - a_1 T^{3/2} - a_2 T^{5/2} - a_3 T^{7/2}\right), \qquad (2)$$

where $T$ is temperature, and $a_0$, $a_1$, $a_2$, and $a_3$ are positive coefficients.

As shown in Fig. 2(b), the extracted FMR field linewidths were fitted with the linear equation[20] $\Delta H = \Delta H_0 + 4\pi\alpha f/\gamma$, where $\Delta H$ is the field linewidth and $\Delta H_0$ is the extrinsic field linewidth. This enabled the extraction of the intrinsic Gilbert damping ($\alpha$) from the fit line slopes. Figure 2(c) shows that $\alpha$ increases as the temperature decreases. The value of $\alpha$ at room temperature was found to be 0.0025, which is comparable with the previously reported room temperature value for CFAS.[8] The trend of $\alpha$ with temperature is consistent with previous first-principle calculations,[9] and could be attributed to longer electron scattering time at lower temperatures, due to a reduction in phonon-electron scattering. Consequently, the angular momentum transfer at low temperatures occurs predominantly by direct damping through intraband transitions.[11] Similar temperature-dependence of $\alpha$ has also been observed experimentally. For example, the α of $Co_{20}Fe_{60}B_{20}$ has been found to increase by a factor of 3 from 0.007 at 300 K, to 0.023 at 5 K.[11] This is comparable to our results, where α increases by a factor of almost 6 from 0.0025 at 300 K to 0.014 at 13 K. It should be noted that spin pumping into the Au cap layer could have contributed to the measured resonance linewidth, thus causing the extracted α to be higher than its actual value ($α_{CFAS}$). Thus, α = $α_{CFAS}$ + $α_{sp}$, where $α_{sp}$ denotes the spin pumping contribution to the damping.[21] While an investigation of $α_{sp}$ in the CFAS/Au system would exceed the scope of this work, $α_{sp}$ values for a Fe/Au system[21,22] have nonetheless been included in Fig. 2(c) to provide a gauge of the temperature dependence of $α_{sp}$, as well as a rough estimation of the magnitude of $α_{sp}$ in the CFAS/Au system. Figure 2(d) shows an increase in $ΔH_0$ as temperature increases. This could be due to the effect of temperature on the interaction between magnetic precession and sample inhomogeneities, or on magnon-magnon scattering, as these factors contribute to $ΔH_0$.[20,23] In both Fig. 2(c) and 2(d), room temperature values of α and $ΔH_0$ for sputter-deposited CFAS were included, for comparison with the MBE sample. It can be seen that the α is higher for the sputter-deposited sample, consistent with lower half-metallic character due to greater structural disorder.[1,6]

We have also measured the time domain PIMM data at 300 K as shown in Fig. 3(a), where SW15 and SW30 denote edge-to-edge signal line separations of 15 and 30 μm, respectively. The width of all the signal lines was fixed at 10 μm. Using the temporal positions of the centers of the Gaussian wavepackets ($t_{15}$ and $t_{30}$, respectively), the group velocity ($v_g$) was calculated with the equation[5,24] $v_g$ = 15 μm/($t_{30}$ − $t_{15}$). Fast Fourier transform (FFT) was performed on the PIMM data, as shown in Fig. 3(b), verifying the presence of multiple modes, where each mode manifested as a dark-light-dark oscillation. The $v_g$ decreases from 26 km/s at 50 Oe to 11 km/s at 370 Oe, as shown in Fig. 3(c). Moreover, from the VNA transmission data, which is another measure of spin wave propagation, attenuation length ($\Lambda$) and $\alpha$ were extracted as a function of magnetic field at room temperature, using the method reported elsewhere.[24,25] The spin wave amplitude was extracted from Lorentzian fittings of the VNA transmission resonance peaks, which were measured using waveguides with different center-to-center signal line-signal line (S-S) spacings. Then, $\Lambda$ was extracted using the equation[24] $A_1\exp(x_1/\Lambda) = A_2\exp(x_2/\Lambda)$, where $A_1$ and $A_2$ denote the measured spin wave amplitudes, while $x_1$ and $x_2$ denote the different S-S spacings for the corresponding waveguides. The $\Lambda$ decreases from 23.3 µm at 460 Oe, to 12.1 µm at 1430 Oe, as shown in Fig. 3(c). Using the following equation,[25] $\alpha$ was calculated at different magnetic fields, as shown in Fig. 3(d)

$$\alpha = \frac{\gamma(2\pi M_s)^2 d e^{-2kd}}{2\pi f \Lambda (H + 2\pi M_s)} \quad (3)$$

where $d$ is the film thickness and $k$ is a spin wave vector, which can be estimated by $2\pi/$(signal line width).[5] The $\alpha$ values (0.0026 – 0.0031) are consistent with the room temperature value (0.0025) obtained from the FMR measurements.

As shown in Fig. 3(c), $\Lambda$ and $v_g$ decreased as the applied magnetic field increased, consistent with previous experimental[11] and theoretical[5] results. This trend can be understood in terms of the following equation,[5,26]

$$v_g = \frac{\gamma^2 \mu_0^2 M_s^2 d}{8\pi f} e^{-2kd}, \quad (4)$$

where $\mu_0$ is the permeability of free space. As the applied magnetic field increases, the resonance frequency increases, thus $v_g$ decreases. In addition, for a given value of $\alpha$, the magnetic precession will decay within a certain amount of time. Hence, the distance travelled by the precessional disturbance within that amount of time depends on its propagation velocity, $v_g$. Consequently, the higher the $v_g$, the longer the distance travelled, and thus, the higher the $\Lambda$. The obtained values of $\Lambda$ and $v_g$ are comparable to those of other ferromagnetic materials for the Damon-Eshbach surface spin wave mode.[5,11,12] For example, $\Lambda$ of 18.95 μm was extracted for CFA by micromagnetic simulations,[5] while $\Lambda$ and $v_g$ values as high as 23.9 μm and 25 km/s, respectively, were experimentally observed in CoFeB.[11] Furthermore, $\Lambda$ as high as 16.7 μm was experimentally observed in CMFS.[12]

In conclusion, we have found a decreasing trend of $\alpha$ with increasing temperature for MBE-grown $Co_2FeAl_{0.5}Si_{0.5}$, in the temperature range of 13 – 300 K. The room temperature value of $\alpha$ was found to be 0.0025, which was approximately 6 times lower than that at 13 K. We have also investigated $v_g$ and $\Lambda$ in CFAS, obtaining values as high as 26 km/s and 23.3 μm respectively, at room temperature.

This work was supported by the Singapore NRF CRP Award No. NRF-CRP 4-2008-06.

**Figure captions**

FIG. 1. (a) Normalized magnetic hysteresis data along the crystallographic hard [100] and easy [110] axes of MBE-grown CFAS. $M_s$ is the saturation magnetization. (b) θ-2θ XRD data of the MBE-grown CFAS sample. (c) Optical microscopy image of the CFAS mesa integrated with asymmetric coplanar waveguides (ACPW). The orientation of the in-plane magnetic field ($H$) is indicated.

FIG. 2. (a) FMR frequency at different magnetic fields. Inset: FMR frequency for a fixed field (1040 Oe) at different temperatures. (b) Resonance linewidth as a function of frequency at different temperatures (symbols), with corresponding fit lines. (c) Gilbert damping parameter ($\alpha$) at different temperatures. The spin pumping contribution to damping ($\alpha_{sp}$) for a Fe/Au system has been included, where all $\alpha_{sp}$ values were obtained from literature, except those at 13 K and room temperature, which were obtained by extrapolating the literature values. (d) Extrinsic field linewidth ($\Delta H_0$) at different temperatures.

FIG. 3. (a) PIMM data from two different signal line-signal line spacings at $H = 50$ Oe for 300 K. (b) Fast Fourier transform (FFT) of room temperature PIMM data. (c) Room temperature group velocity ($v_g$, axis: left and bottom) and attenuation length ($\Lambda$, axis: top and right) at different magnetic fields. (d) Room temperature Gilbert damping parameter ($\alpha$) at different magnetic fields.

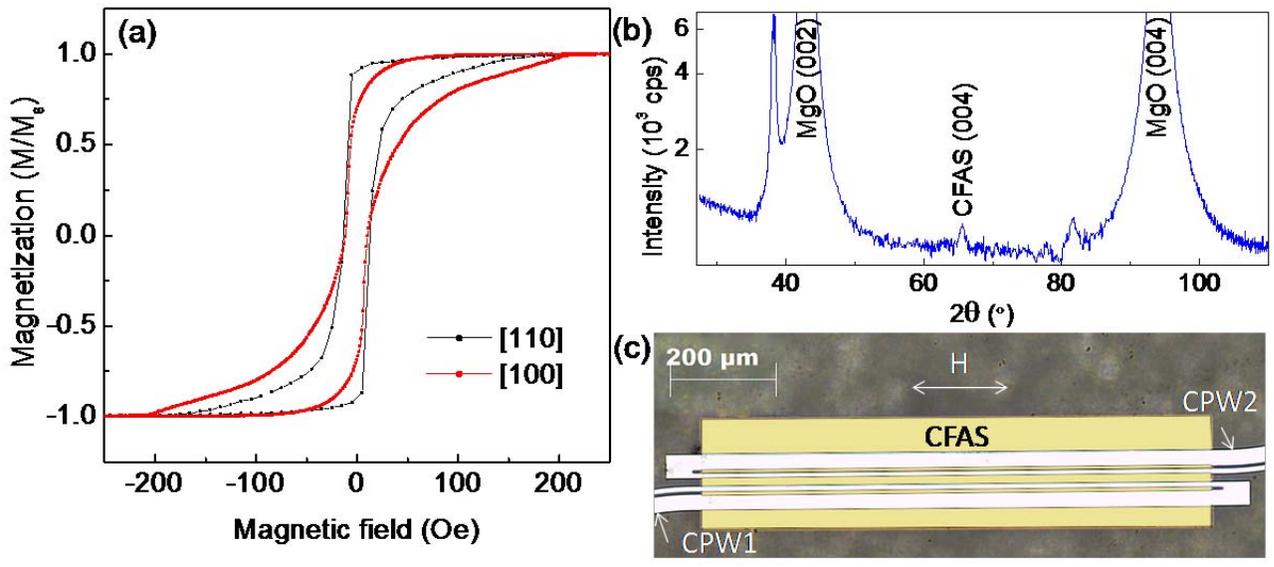

FIG. 1

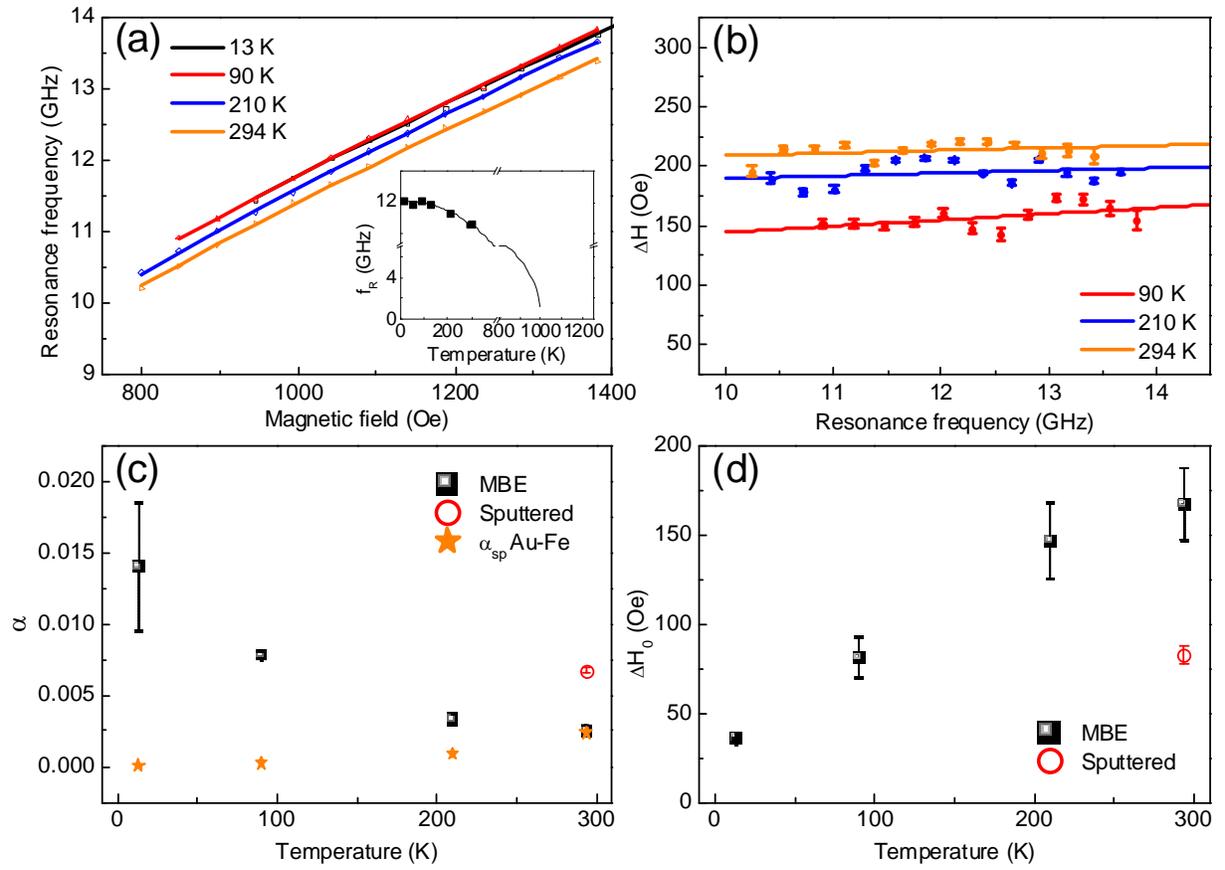

FIG. 2

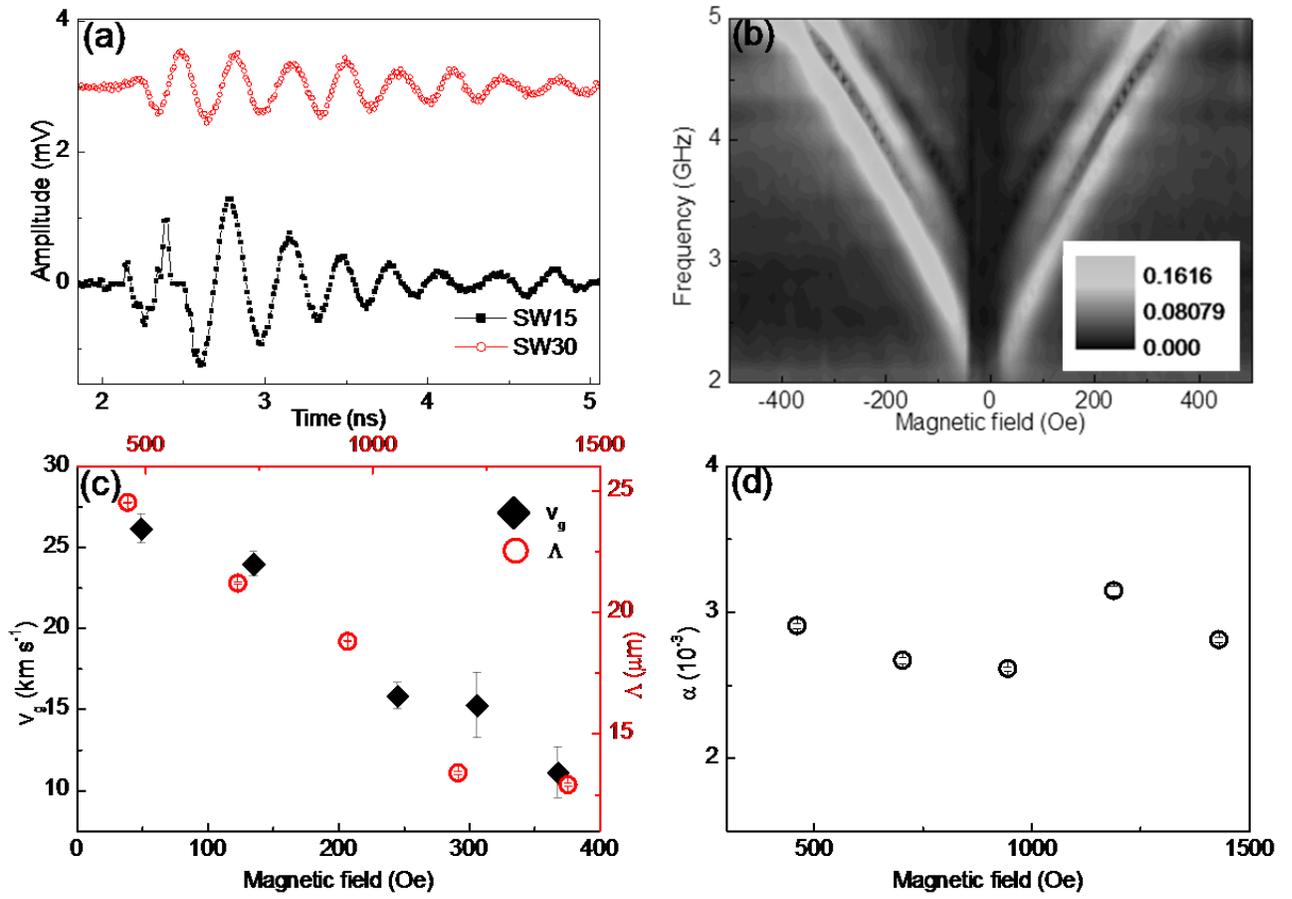

FIG. 3